# Analysis of the satellite navigational data in the Baseband signal processing of Galileo E5 AltBOC signal

Subhan Khan[1*], Yiqun Zhu[2], Muhammad Jawad[1], Muhammad Umair Safder[1], Mujtaba Hussain Jaffery[1], Salman Javid[1]

[1]Department of Electrical Engineering, COMSATS Institute of Information Technology , Lahore, Pakistan
[2]Department of Electronics & Electrical Engineering, University of Nottingham, Nottingham, UK.
Corresponding author: *subhankhan@cuilahore.edu.pk

**Abstract:** The advancement in the world of global satellite systems has enhanced the accuracy and reliability between various constellations. However, these enhancements have made numerous challenges for the receivers' designer. For instance, comparing the acquisition and tracking of the Galileo signals turn into relatively complex processes after the utilization of Alternate-Binary Offset Carrier (AltBOC) modulation scheme. This paper presents an efficient and unique method for comparing baseband signal processing of the complex receiver structure of the Galileo E5 AltBOC signal. More specifically, the data demodulation has attained after comparing the noisy satellite data sets with the clean data sets. Moreover, the paper presents the implementation of signal acquisition, code tracking, multipath noise characteristics, and carrier tracking for various datasets of satellite after the eradication of noise. The results obtained in the paper are promising and provide the through treatment to the problem.

**Keywords:** Carrier tracking, code tracking, Galileo E5AltBOC, multipath noise characteristics, signal acquisition

## 1 INTRODUCTION

Galileo is European's Global Navigation Satellite System (GNSS), which is currently at the development stage. The satellites of GNSS constantly transmit signals in two or more frequencies in L band. The signals have two parts, such as ranging codes and navigation date, which enables the user to calculate signal travelling time from satellite to receiver and satellite's coordinates at each epoch time. The GNSS mainly consist of E1, E5 and E6 band signals. The most sophisticated Galileo E5 signal is further split into E5a and E5b band signals and employs AltBOC modulation that can be used as a wideband signal to offer an accuracy of 20 centimeters even with pseudo-range only. The AltBOC is a complex signal composed of four codes multiplexed to have a constant envelope. The main lobes of the signal span over 50 MHz range that means the signal bandwidth is about thirty times larger than the current GPS signal's bandwidth and the signal uses complex modulation [2]. However, the intricacy of the E5 AltBOC signal is extremely challenging for the baseband signal processing. The AltBOC consist of four Pseudo-Random Noise (PRN) codes that are used for the transmission of four channels, such as E5a-I, E5a-Q, E5b-I, and E5b-Q. The E5a-I and E5b-I (the in-phase components) carry the data; therefore, these two channels are called data channels. Moreover, the E5a-Q and E5b-Q (the quadrature components) do not carry any data; therefore, these two channels are called pilot channels.

The first stage in the baseband signal processing is the signal acquisition that gives the rough estimation of the Doppler frequency and code delay of the received signal [3]. Tracking is the second stage in the baseband signal processing of Galileo receiver that refines coarse values obtained from the signal acquisition stage. Moreover, the second stage also keeps track of the signal properties that may change over time to demodulate the navigation data from a specific satellite [4]. The navigation data extraction is the third stage in the baseband signal processing section of a Galileo receiver. When the signals are correctly tracked, then the carrier wave and the PRN code can be eliminated from the signal to leave only the navigation bits with a period of 20*ms*.

The main contributions of the paper are stated as:

- A detailed baseband signal processing analysis of the Galileo E5 AltBOC.
- To compare raw satellite data, noisy data, and filtered data for the positioning in the presence of software based receiver.
- To distinguish the auto-correlation functions and signal tracking using the satellite navigational data.

The software application-based learning of Galileo E5 AltBOC signal is still the area to be look upon; however, several research articles have discussed the baseband signal processing of Galileo E5 signal. In [5], the authors presented the signal acquisition of the E5 Galileo signals using parallel frequency search algorithm developed in MATLAB. A code was developed to construct the reconfigurable receiver for a wide range of applications [6]. The architecture of the receiver was designed on software-defined radio (SDR) techniques [6]. However, the authors were unable to perform the signal tracking. In [7], the basis of baseband signal processing is presented but without discussing tracking loop parameters that are important in the development of code and carrier loops. An innovative and efficient solution is proposed for the demodulation of the navigation data [8]. However, the authors did not analyze the performance of the tracking stages in delay locked loops under hazardous situation.

In [9], a novel approach is addressed to perform full band operation along with wiping off the data bias on both E5a and E5b signals to take maximum advantage of the power. The introduction of mitigation in the code phase multipath by frequency exploitation is used for E5 signal [9]. Another demodulation approach of the Side Band Translator (SBT) is proposed in [10]. The authors have recovered the data channels by frequency shift of the real signals on the received signal and obtained replicas from the sub-carrier [10]. Once the data is acquired then both frequency shifts are filtered in a separate way to reduce cross correlations [10]. The interferences of channels were not presented in Ref. [10]. The need of the interferences with adjacent channels are performed in [11]. The navigation data is extracted and separately acquired in the two BPSK signals by performing demodulated operation. However, there is no signal acquisition performed by the authors. A complete Galileo E5 AltBOC (15, 10) receiver is designed by [12-14]. The authors present fundamentals of acquisition, tracking, and demodulation of the AltBOC (15, 10). The demodulation and tracking channels are achieved with hardware prototype of the receiver. However, the acquisition algorithm is not efficient and auto-correlation function (ACF) of the structures has less narrow side lobes. The high-accuracy positioning achieved by the multiple global GNSS has an advantageous impact to eradicate systematic errors. The approach based on semiparametric estimation for eliminating systematic errors has been experimental proved in [16]. In terms of performance of the E5 data, the multipath effect has instantaneously resolve the current constellation data propagation [17].

In addition to the aforementioned research, many authors have focused on engineering education, such as [18]-[23]; however, to the best of our knowledge, no prior work exist on the engineering education tool for the baseband signal processing of Galileo E5 signal. In terms of carrier tracking, the new generation signals opt a robust multiple update-rate Kalman filter (MUKF). The robustness and improvement in the generalized model of MUKF is compared in the simulated and experimental environment which shows the better performance of robust MUKF over dual-phase locked loop [24]. In [25], a methodology is proposed which computes the same adjustment for the satellite and differential code biases. The study concludes the differential code biases and vertical total electron content are closely correlated. The main theme of this research is to compute GPS P1-P2 DCBs with Jason-2 spacecraft.

The accuracy and recent trends in Galileo are presented in [26]. The stability in clock, frequency, and positioning are the key factors for a navigation system. Galileo provides a precision of 2cm in static mode while keeping the daily solutions. Recently, various GNSS are compared in terms of single frequency (L5/E5a) when the multi-GNSS observations are different with respect to a common pivot satellite [27]. The study elaborated estimation and applications of inter-system biases. However, the status of Galileo's clock, frequency, and positioning are neglected. In terms of high sensitivity acquisition of Galileo, the secondary codes are preferred to attain better cross-correlations to synchronize the satellite data to help interference elimination. The emergence of latest modulations schemes reduces the impact of multipath and allow higher chipping rates to get better accuracy [28]. Despite of many latest research available in the field of GNSS, the comparison of satellite data in the presence of noise and filter are not discussed. This paper not only presents the comparison of clean satellite data with noisy and filtered data but also a detailed analysis in the modeling of baseband signal processing of Galileo E5 AltBOC signal.

The paper is organized as follows: Section 2 explains Galileo E5 signal structure. The baseband signal processing in the form of acquisition and tracking is presented in Section 3 and Section 4, respectively. The results are discussed in section 5. Section 6 concludes the paper along with the future directions.

## 2 GALILEO E5 SIGNAL STRUCTURE

The Galileo E5 signal is a Right Hand Circular Polarized (RHCP) signal. The modulation scheme used to modulate this signal is called Alternate Binary Offset Carrier (Alt-BOC) modulation. Modulation of carrier is performed by four PRN codes which are quasi-orthogonal to each other. These codes are $c_{E5a\text{-}I}$, $c_{E5a\text{-}Q}$, $c_{E5b\text{-}I}$, and $c_{E5b\text{-}Q}$. Apart from the PRN codes, two navigation messages ($d_{E5a\text{-}I}$ and $d_{E5b\text{-}I}$) and one side band sub-carrier also take part in the modulation. Following expressions can be derived for the bandpass transmitted signal by assuming one satellite at a time.

$$S_{E5_t}(t) = AR[s_{E5}(t)e^{j2\pi f_c t}] \tag{1}$$

$$S_{E5_t}(t) = s_{E5-I}(t) + js_{E5-Q}(t) \tag{2}$$

where $A$ is Amplitude of the signal, $f_s$ is the Carrier frequency that is selected as 1191.795 *MHz*, and $R$ is the Real function. The generation of Alt-BOC modulated Galileo E5 signal has been described in European Space Agency document [24]:

$$\begin{aligned} s_{E5}(t) = & \frac{1}{2\sqrt{2}}\left(e_{E5a-I}(t) + je_{E5a-Q}(t)\right)[sc_{E5-s}(t) - jsc_{E5-s}\,(t - \frac{T_{s,E5}}{4}) \\ & + \frac{1}{2\sqrt{2}}\left(e_{E5b-I}(t) + je_{E5b-Q}(t)\right)[sc_{E5-s}(t) + jsc_{E5-s}\,(t - \frac{T_{s,E5}}{4}) \\ & + \frac{1}{2\sqrt{2}}\left(\bar{e}_{\,E5a-I}(t) + j\bar{e}_{\,E5a-Q}(t)\right)[sc_{E5-p}(t) - jsc_{E5-p}\,(t - \frac{T_{s,E5}}{4}) \\ & + \frac{1}{2\sqrt{2}}\left(\bar{e}_{\,E5b-I}(t) + j\bar{e}_{\,E5b-Q}(t)\right)[sc_{E5-p}(t) + jsc_{E5-p}\,(t - \frac{T_{s,E5}}{4}) \end{aligned} \quad (3)$$

In the eq 3, the signal components $e_{E5a\text{-}I}$, $e_{E5a\text{-}Q}$, $e_{E5b\text{-}I}$, and $e_{E5b\text{-}Q}$ carry the PRN codes as well as the navigation messages. Whereas, the dashed components $\bar{e}_{E5a\text{-}I}$, $\bar{e}_{E5a\text{-}Q}$, $\bar{e}_{E5b\text{-}I}$, and $\bar{e}_{E5b\text{-}Q}$ denote the product signals. The symbols $sc_{E5-s}$ and $sc_{E5-p}$ denote the four-valued sub-carrier functions for the single signals side bands (SSB) and the product signals side bands (PSB). The spectrum is split into two side lobes which are completely symmetric due to the existence of sub-carrier signals. These two side lobes are called as E5a and E5b and they are centered at $\pm f_{sc}$ MHz from the carrier frequency. The following expression represents its envelope:

$$G_{AltBOC}(f) = \frac{4f_p}{\pi^2 f^2}\frac{cos^2\left(\frac{\pi f}{f_p}\right)}{a^2}\left[a^2 - a - 2a\cos\left(\frac{\pi f}{4f_{sc}}\right) + 2\right] \quad (4)$$

where $f_p$ is the PRN code with frequency as 10.23 MHz and $a = \cos(\frac{\pi f}{2f_{sc}})$. In GNSS bands, the power spectral density (PDF) of the E5 Alt-BOC (15,10) signal has one of the broadest spectrum. The satellite has a transmission bandwidth of 90MHz. The minimum required value of the bandwidth for the receiver is 51.15 MHz. The continuous envelope of the modulated signal is extracted through the product signals The PSB's first harmonic is found at -3$f_{sc}$ (-45 MHz) and it holds 61.5% of the total power allocated for the product sub-carriers that is also 14.64% of the total E5 signal power. Therefore, the 0.615*0.1464 = 9% of the total power is hold by the harmonics at $\pm 3f_{sc}$. The first harmonic of the SSB is located at $+f_{sc}$ (15 MHz) from the center frequency and it holds 94.96% of the total power allocated for the single sub-carriers that is 85.36% of the total power of the E5 signal. The two main lobes of the Alt-BOC signal (E5a and E5b) posses 0.9496 * 0.8536 = 81% of the total power. As most of their power is filtered out, the Eqn. 3 is rewritten as:

$$\begin{aligned} s_{E5}(t) \approx & \frac{1}{2\sqrt{2}}\left(e_{E5a-I}(t) + je_{E5a-Q}(t)\right)[sc_{E5-s}(t) - jsc_{E5-s}\,(t - \frac{T_{s,E5}}{4}) \\ & + \frac{1}{2\sqrt{2}}\left(e_{E5b-I}(t) + je_{E5b-Q}(t)\right)[sc_{E5-s}(t) + jsc_{E5-s}\,(t - \frac{T_{s,E5}}{4}) \end{aligned} \quad (5)$$

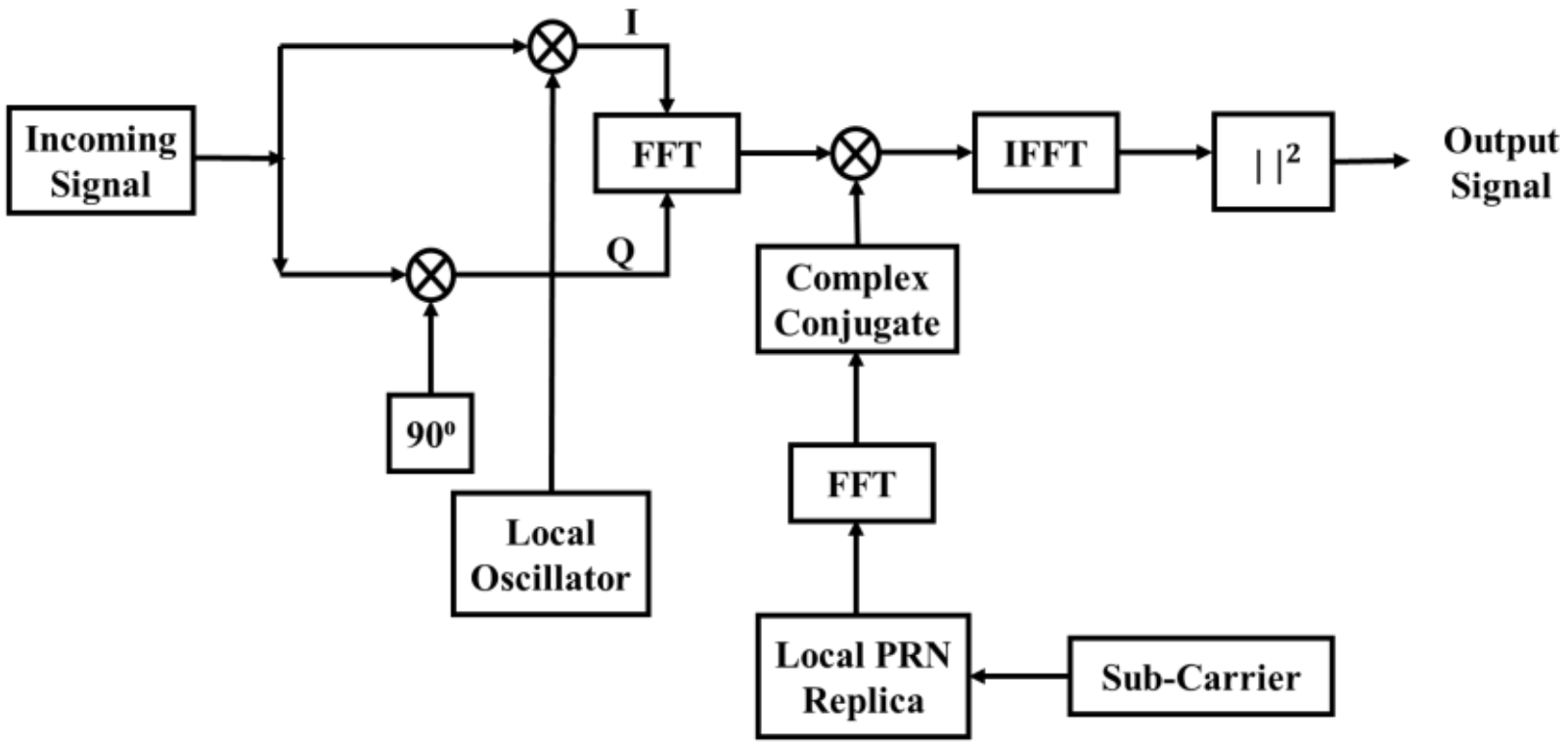

Figure 1: Block diagram of signal acquisition

Similarly, using a band pass filter having a cut-off frequency less than 90 MHz, the multiplication results of intermodulation at $\pm 3f_{sc}$ and $\pm 5f_{sc}$ frequencies are mainly eliminated. Therefore, the single subcarrier expressions are also approximated as a pure sine and cosine functions, as shown in Eqn. 6 and Eqn. 7.

$$sc_{E5-s}(t) \approx \cos(2\pi f_{sc} t) \tag{6}$$

$$sc_{E5-s}\left(t - \frac{T_{s,E5}}{4}\right) \approx \sin(2\pi f_{sc} t) \tag{7}$$

Using Eqn. 6 and Eqn. 7, the expression for Eqn. 5 is rewritten as:

$$s_{E5}(t) \approx \frac{1}{2\sqrt{2}}\left[E5a(t)e^{-j2\pi f_{sc}t} + E5b(t)e^{j2\pi f_{sc}t}\right] \tag{8}$$

$$\text{where } E5a(t) = e_{E5a-I}(t) + je_{E5a-Q}(t) \tag{9}$$

$$\text{and } E5b(t) = e_{E5b-I}(t) + je_{E5b-Q}(t) \tag{10}$$

# 3 SIGNAL ACQUISITION

The two pilot channels E5a-Q and E5b-Q are used for the analysis of the ACF. Because both pilot channels do not carry any data; therefore, a more robust tracking is obtained by using a relatively larger integration time. By using the imaginary parts alone in Eq 9 and 10, it can be written as:

$$s_Q(t) = \tilde{c}_{\,E5a-Q}(t-\tau)e^{-j(2\pi f_{sc}(t-\tau))} + \tilde{c}_{\,E5b-Q}(t-\tau)e^{j(2\pi f_{sc}(t-\tau))} \tag{11}$$

The expression $S_Q$ should be correlated with $c_{E5a\text{-}Q}$ and $c_{E5b\text{-}Q}$ multiplied with the complex conjugate of the corresponding sub-carrier exponential in order to track the pilot component.

$$\tilde{R}_{\,Q}(\tau) = \tilde{R}_{\,E5a-Q}(\tau) + \tilde{R}_{\,E5b-Q}(\tau) \tag{12}$$

$$\tilde{R}_{E5a-Q}(\tau) = \int_{T_{int}} \tilde{c}_{E5a-Q}(t-\tau)e^{-j(2\pi f_{sc}(t-\tau))}\, c_{E5a-Q}(t)e^{j2\pi f_{sc}(t)} \approx \tilde{R}(\tau)e^{-j2\pi f_{sc}\tau} \tag{13}$$

$$\tilde{R}_{E5b-Q}(\tau) = \int_{T_{int}} \tilde{c}_{E5b-Q}(t-\tau)e^{-j(2\pi f_{sc}(t-\tau))}\, c_{E5b-Q}(t)e^{-j2\pi f_{sc}(t)} \approx \tilde{R}(\tau)e^{j2\pi f_{sc}\tau} \tag{14}$$

where $\tilde{R}(\tau)$ is the triangular function, and the integration time is denoted by $T_{int}$.

Figure 1 illustrates the signal acquisition processes, where the arriving signal is split into in-phase and quadrature components. Both components are multiplied with the local oscillator and a 90° phase shift is added to the local oscillator before being multiplied with the quadrature component. The Local oscillator multiplies different carrier frequencies with the aim to eliminate the carrier frequency of the arriving signal. Both components are eventually combined and transformed into frequency domain using the Fourier method. The tracking output is then multiplied with the Fourier-transformed complex conjugate of code phases to remove code phase from the arriving signal. Finally, an inverse Fourier transform is applied to achieve the acquisition result.

# 4 SIGNAL TRACKING

Carrier loop and the PRN code tracking loop constitute the fundamentals of tracking in order to refine the coarse values of carrier frequency and code phase obtained from the acquisition phase and to demodulate the data bits from the signal. The carrier wave tracking is usually performed using Frequency Lock Loop (FLL) or Phase Lock Loop (PLL). Conversely, a Delay Lock Loop (DLL) is used for the signal's code tracking. In GNSS receiver, the code tracking is very important because it gives the pseudo-range measurements and swift code phases for the PLL. Therefore, a DLL's design must be robust and accurate.

Initial step in the tracking is to remove the carrier by multiplying with a perfectly allied carrier replica. Secondly, the resultant in phase and quadrature arms are multiplied with the three diverse copies of the code named as Early, Prompt, and Late. Third, the outputs of the second step are filtered using integration and dump filter sequentially. The output of the integrations shows the correlation of the code replica with the code present in the incoming signal. Figure 2 briefly illustrates the aforementioned basic code tracking process. The correlation depends on the phase of the local carrier wave. If the local carrier wave and the received signals are in phase, then in-phase component will contain the energy. However, if both the components are not in phase, then the energy keeps switching between the in-phase and the quadrature arm.

## 4.1 DLL Discriminators

The DLL discriminator estimates the error in the code delay between the incoming code and the locally created spreading code. Different GNSS receivers use different kinds of discriminators; however, the key characterization between the discriminators is based on whether the discriminators are coherent or non-coherent. The Early Minus Late (EML) is the widely used type of coherent discriminator and is given as: $D_{EML} = I_E - I_L$. The EML discriminator is characterized on the simplicity and its linear response. The disadvantages associated with the EML are (a) the need of a good tracking loop for optimal performance and (b) the EML is sensitive to carrier phase errors. Therefore, the EML is sensitive to $180^0$ phase shifts.

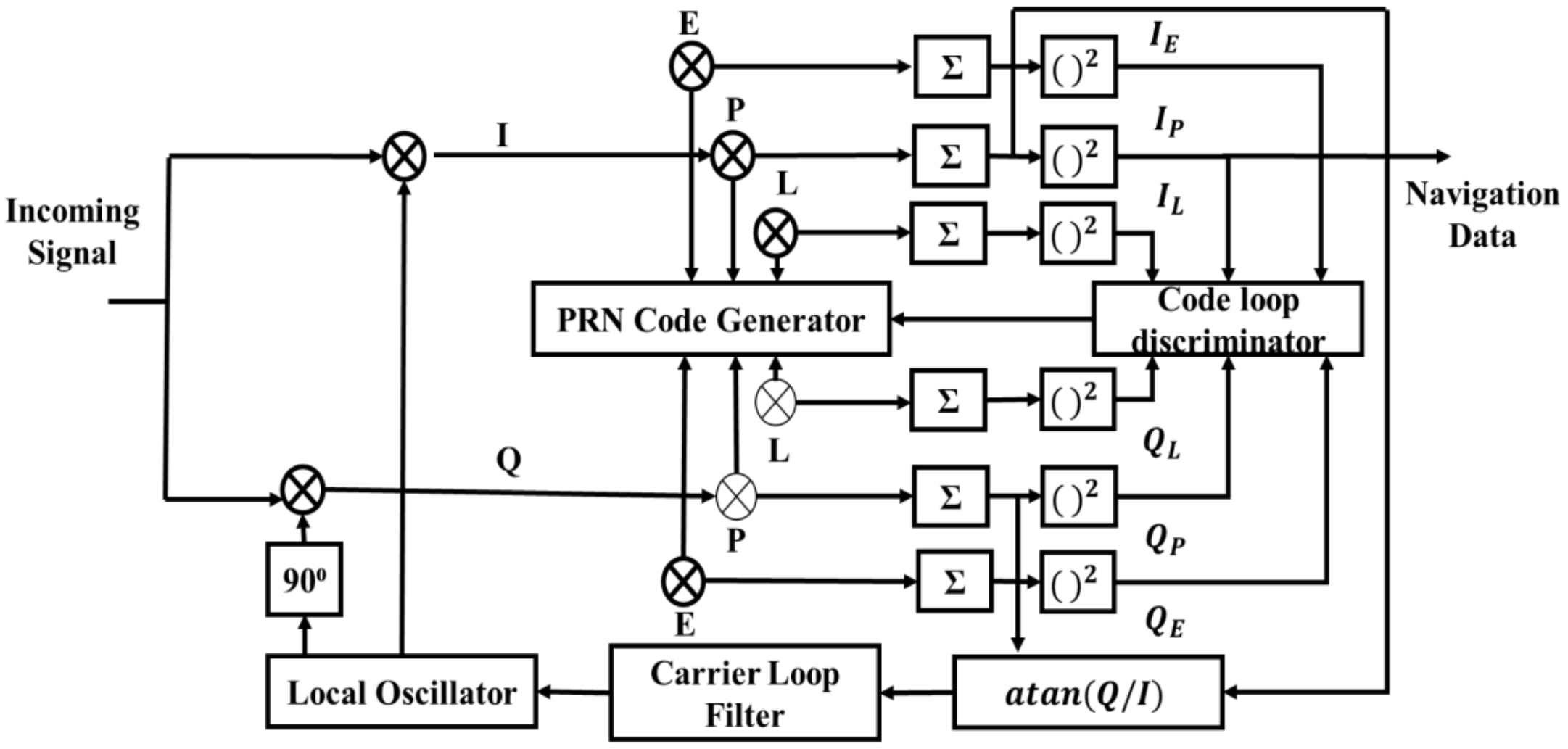

Figure 2: Basic code tracking block diagram

Non-coherent discriminators are frequently used for robust DLL tracking because they are sensitive to the phase errors. Figure 3 illustrates the block diagram of the DLL block with six correlators instead of three correlators. The six correlators have an advantage to eliminate the problem of insensitivity to 180° phase shifts. There are mainly two types of the non-coherent discriminators, such as (a) non-coherent early minus late power (NEMLP) and (b) Dot-Product (DP), as shown in Eqn. 15 and Eqn. 16.

$$D_{NEMLP} = (I_E^2 - I_L^2) - (Q_E^2 - Q_L^2) \quad (15)$$

$$D_{DP} = (I_E - I_L)I_P + (Q_E - Q_L)Q_P \quad (16)$$

The NEMLP discriminator is computational intensive and uses a total of three correlators, such as E, L, and P) whereas the DP discriminator is a less computational intensive and uses only two correlators, such as E-L and P. The NEMLP and DP can be normalized to remove the amplitude sensitivity that in returns improve the performance under a constantly varying SNR. To evaluate the performance of normalization, an experiment is performed with and without using the normalized discriminator and the performance results are compared and analyzed based on stability and linearity. The region near the zero-phase error is named as the stability region, which indicates that for a certain input error, the discriminator will respond in the precise direction. Moreover, the linearity is dependent on the correlator’s spacing and the shape of the ACF.

### 4.2 Code tracking errors

The code tracking loop has three types of errors, such as (a) thermal noise, (b) multipath, and (c) signal dynamics. The reason for dynamic pressure is the satellite-receiver motion that is controlled by the DLL along with PLL and carrier. In this case, the subsequent induced error is very small compared to the other two error sources. If the multipath noise is not present and the signal does not experience any other kind of interference, then the thermal noise is a dominant factor. This noise is a measure of the noise present in the surroundings of the receiver antenna located at the ground station. The mathematical expression given in Eqn. 17 and Eqn. 18 are used to calculate thermal noise jitter for both NEMLP and DP discriminators.

$$\sigma_{NEMLP}^2 = \frac{B_n \int_{-B/2}^{+B/2} G(f) sin^2(\pi f d) df}{\frac{C}{N_o}(2\pi \int_{-B/2}^{+B/2} f G(f) sin(\pi f d) df)^{\ 2}} \beta \tag{17}$$

$$\sigma_{DP}^2 = \frac{B_n \int_{-B/2}^{+B/2} G(f) sin^2(\pi f d) df}{\frac{C}{N_o}(2\pi \int_{-B/2}^{+B/2} f G(f) sin(\pi f d) df)^{\ 2}} \psi \tag{18}$$

Where the $\beta$ and $\psi$ values are given by the following expressions:

$$\beta = (1 + \frac{\int_{-B/2}^{+B/2} G(f) cos^2(\pi f d) df}{T_{int} \frac{C}{N_o}(\int_{-\frac{B}{2}}^{+\frac{B}{2}} G(f)\, cos(\pi f d) df)^{\ 2}} \tag{19}$$

$$\psi = (1 + \frac{1}{T_{int} \frac{C}{N_o} \int_{-B/2}^{+B/2} G(f)} \tag{20}$$

In Eqn. 19 and Eqn. 20, the $G(f)$ is the power spectral density (PSD) of the signal whereas $\frac{C}{N_o}$ represents the noise ratio in the carrier signal. The term $B_n$ is the code loop noise and $B$ is the front-end bandwidth. The Eqn. 19 and Eqn. 20 can be rewritten using the $G(f)$ value from Eqn. 4 as:

$$\sigma_{NEMLP}^2 = \frac{B_n(1 - \tilde{R}_{\ Q}(d))}{2\alpha_{(-d/2)}^2 \frac{C}{N_o}} \left( 1 + \frac{2}{(2 - \alpha_{(-d/2)} d) \frac{C}{N_o} T_{int}} \right) \tag{21}$$

$$\sigma_{DP}^2 = \frac{B_n(1-\tilde{R}_Q(d))}{2\alpha_{(-d/2)}^2 \frac{C}{N_o}}\left(1+\frac{1}{\frac{C}{N_o}T_{int}}\right) \quad (22)$$

Eqn. 21 and Eqn. 22 illustrates the performance of a DP discriminator to be better than NEMLP discriminator in terms of code noise because of the squaring of early and late correlators.

### 4.3 Code Multipath Error

The travelling of signal towards the receiver, the multipath is a phenomenon occurred due to the reflection of the signal from the neighboring objects, such as buildings. These reflected signals reach the receiver with an added delay compared to the signal components that travel in the line of sight. After multiple reflection occurrence, scattering, and other interferences, the signal can be distorted entirely. Because the pseudo-ranges are attained from these measures, the distortions may lead to a false or biased lock and consequently to a range and phase error. The error envelope represents the maximum error present in the signal due to multipath; therefore, it is used to characterize the effect of multipath on the code tracking.

The Galileo E5 signal passes through many phases before it has been fully demodulated, and the data bits have been extracted from it. Tracking phase is one of the critical phases as a lot of sensitive parameters fall in this area. The DLL loop used in the code tracking are affected by parameters, such as (a) front-end filter bandwidth, (b) correlator spacing, (c) discriminator type and (d) even the type of normalization. Different types of discriminators, such as NEMLP and/or DP can be used for different results. Different external parameters, such as receiver's surrounding and multipath also affect the signal that in turn leads to compromising results or it might lead to totally wrong result. Therefore, the code tracking phase of the Galileo E5 signal modulation requires extra attention.

## 5 RESULTS

First, the baseband signal processing operations are performed on the clean data. The clean data is free from all impurities, noises, and losses. Performing the operations on the clean data results in near optimal/ideal output. Second, noise has added in the clean dataset for the purpose of observing the auto-correlation function and the behavior of the tracking signals. Finally, the noisy data has passed through the low-pass filter to eradicate the noise in the data. A simple $8^{th}$ order low-pass filter in the Matlab is sufficient enough to eliminate the noise. The filter is designed in such a way that it can contain the important positing information.

### 5.1 Acquisition under clean data

The Figure 3 illustrates the signal acquisition response in the form of auto-correlation function, code search at 7.8 MHz, frequency search at 7.8 MHz for PRN 17, and the data extraction from the tracking channels without any hazardous or unwanted noise. The datasets for the acquisition are clean ones acquired from the satellite. The ACF of the AltBOC (15, 10) is normalized with side lobes at both ends of the E5A and E5B signals, the samples are the mixture of both data and pilot channel of the E5 signal. The code and frequency search are set at PRN 17 due to the accuracy shown by the clean datasets at 7.8 MHz. The correlation in phase arm and navigation data shows the modulated AltBOC in timing axis that is a proof of acquiring accurate carrier replicas from the data channels.

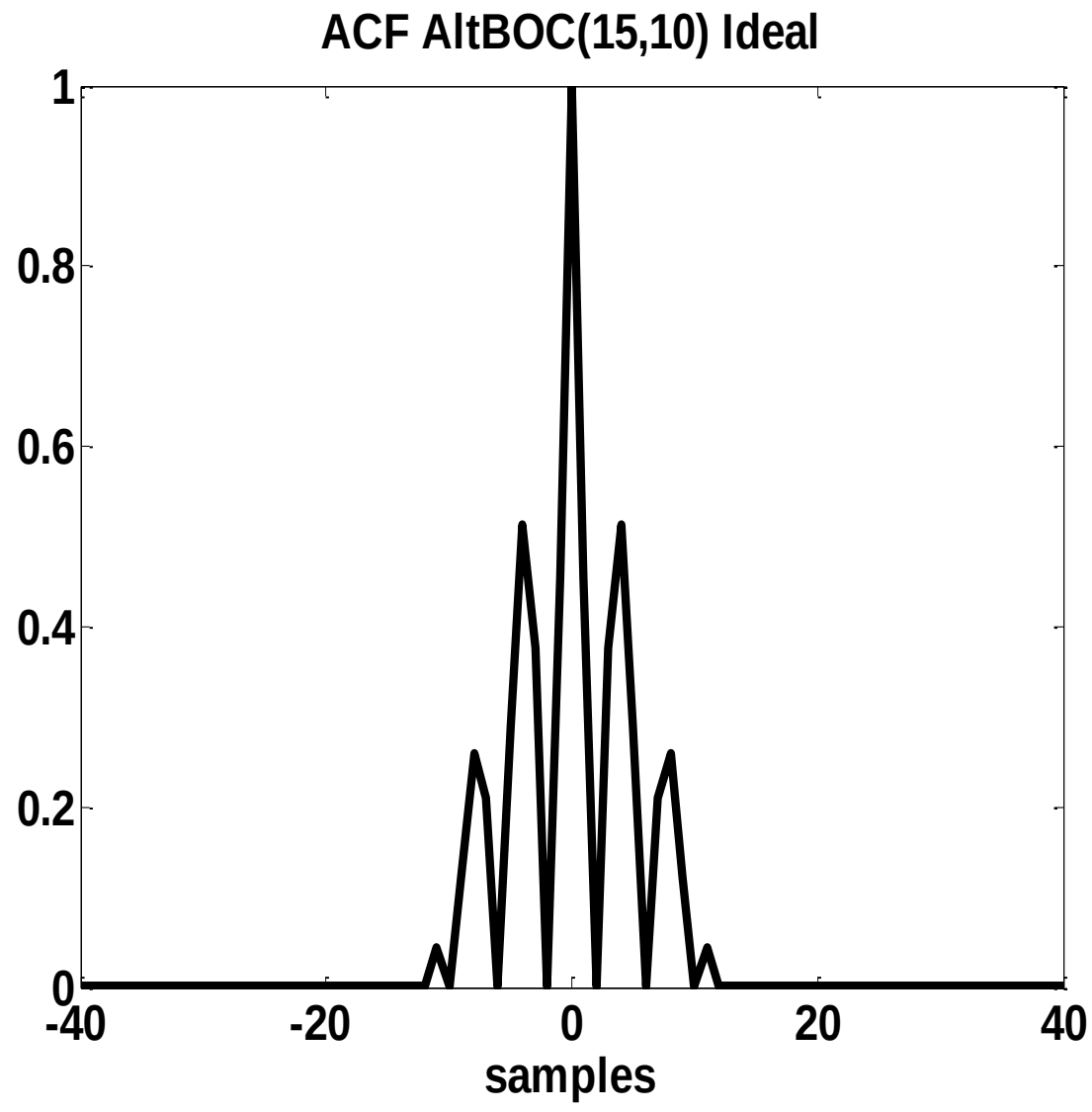

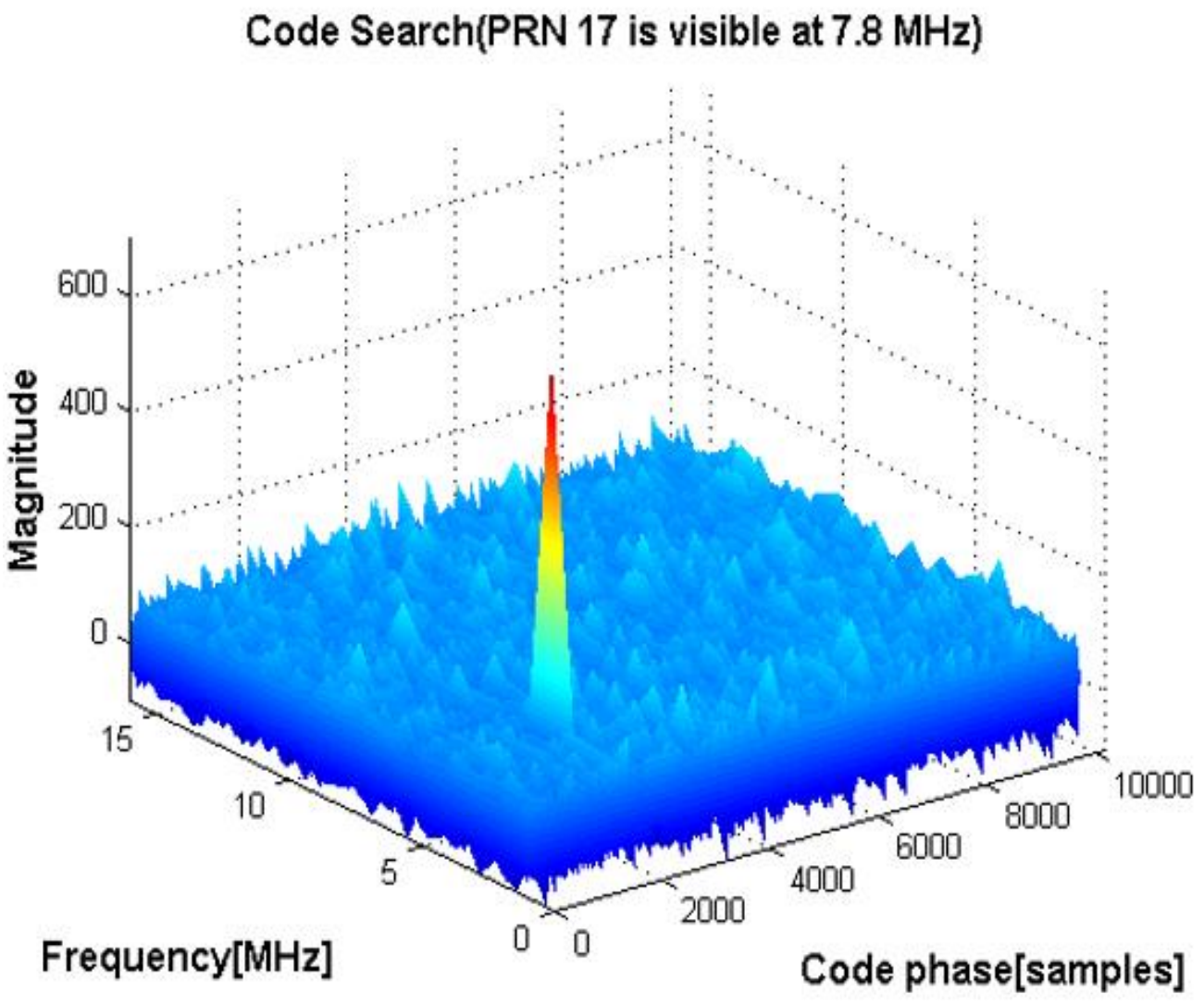

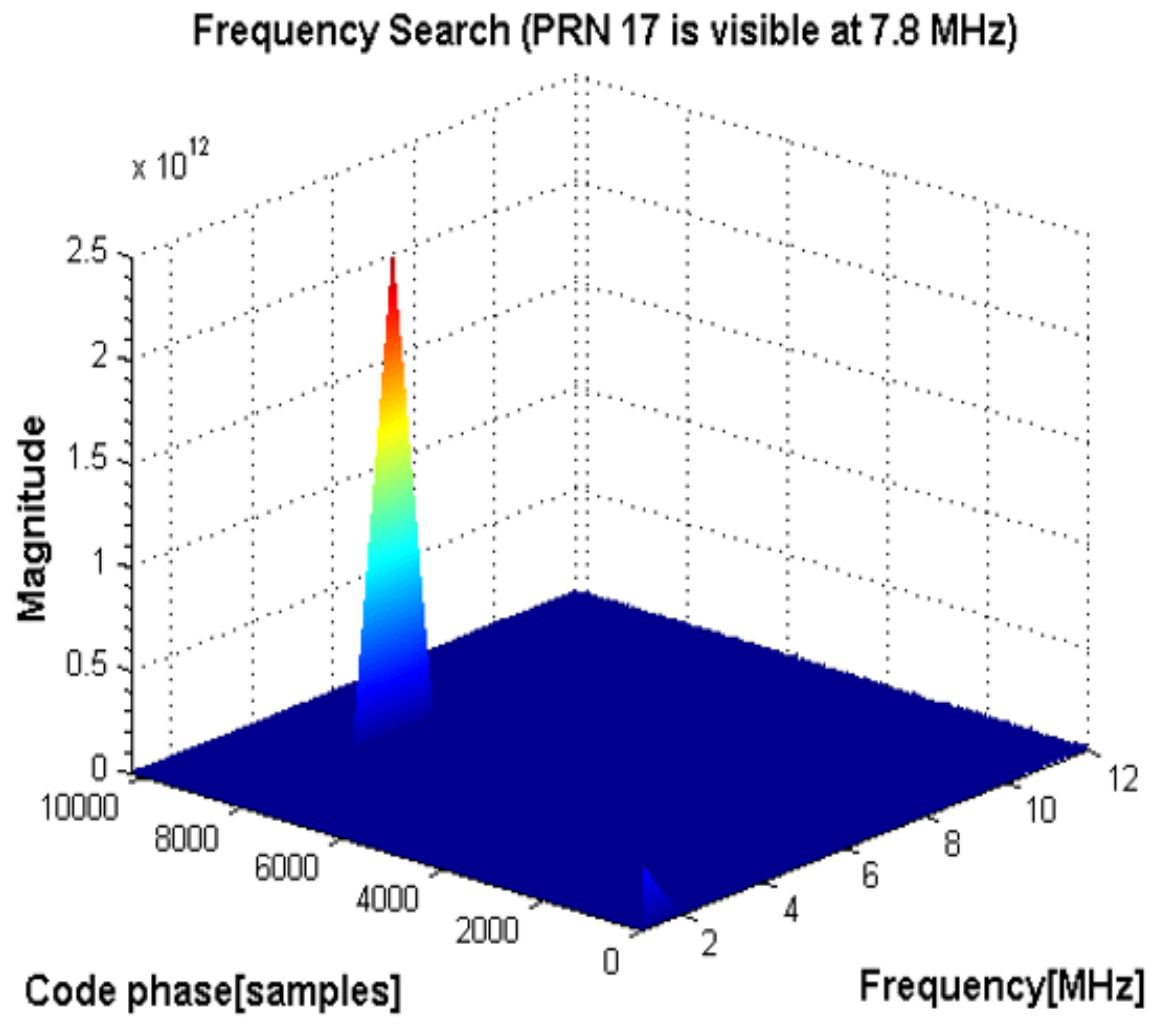

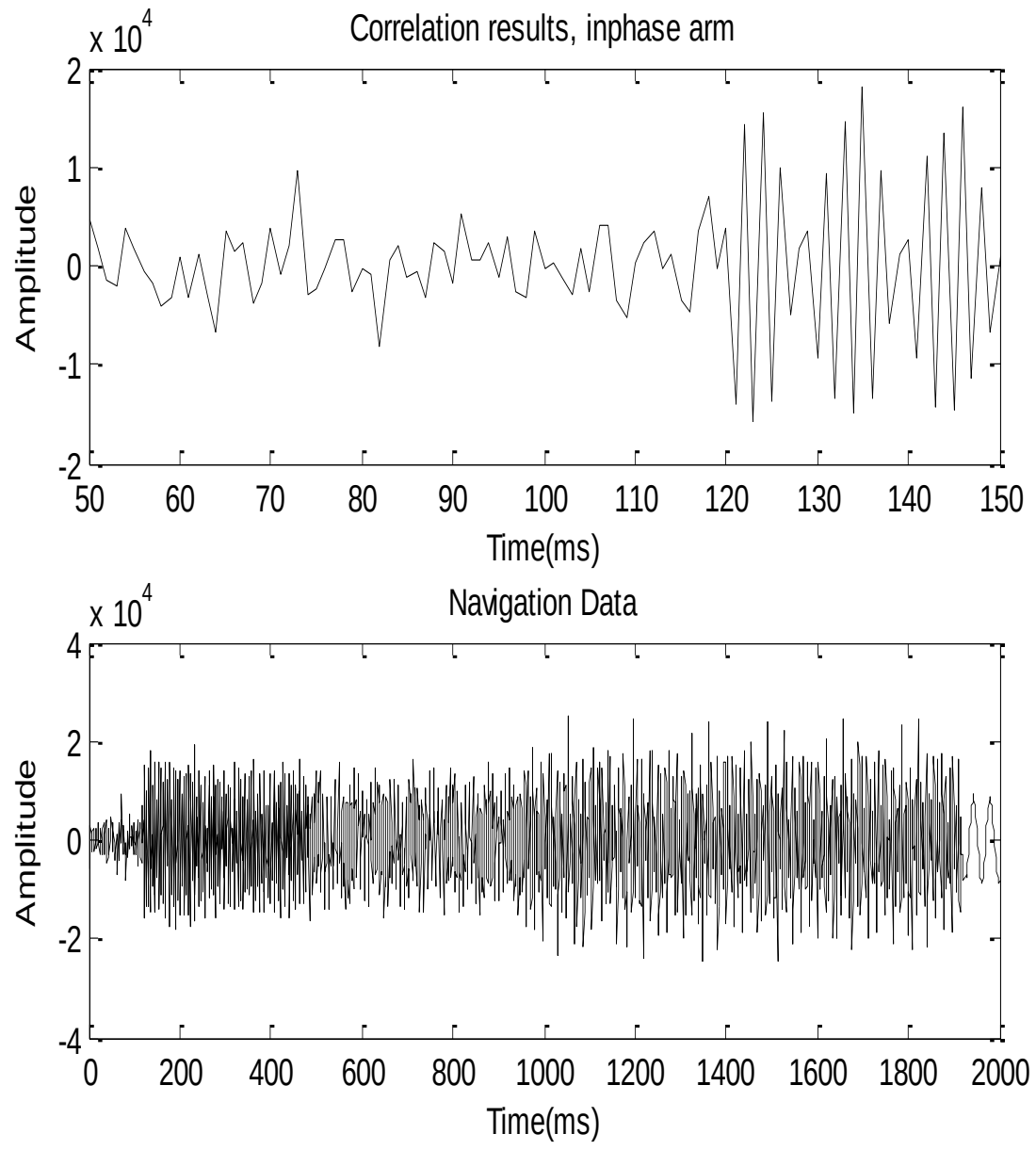

Figure 3: (a) Auto-correlation function of the signal acquisition under clean dataset. (b) Code search using acquisition search algorithm at 7.8 MHz (c) Frequency search using parallel search algorithm (d) Correlation results from the signal tracking after clean datasets

## 5.2 Noisy Data

After performing all the operations on the clean data, the next step is to repeat all the same processes on the noisy data. The noisy data is obtained by multiplying a sum of randomly generated noises with the original data itself. This multiplication results in the addition of noise in the actual data at some random points. This results in the manipulation of the navigational data.

The purpose of all this is to obtain the set of results for the operations performed on the noisy data. Later on, a comparison can be developed between the results obtained by operations performed on clean data and noisy data. This comparison helps in understanding the differences that noise impacts on the data. Acquisition performed on the noisy data results in the following graph as shown in figure 4a. When compared with the acquisition of the clean data, it can be clearly seen that the side lobes tend to disappear with the major lobe occupying the most of the portion indicating the presence of noise.

The three different carrier replicas are generated in order to perform tracking through noisy data sets. The difference, however, lies in the next step where the result of tracking differs due to the fact that a noisy data is being multiplied with those time-delayed carrier replicas. The results obtained indicate a clear presence of noise as they tend to occupy much more space on the graph. Figure 4b shows the navigational data extracted after the introduction of noise.

## 5.3 Filtered Data

The next process is to filter the noisy data in order to try and recover the original data from it. This serves as a model for the practical use, as the data received from the satellite might also experience different kind of noises added to it. The results obtained after performing the operation on the filtered data are not quite similar to the clean data results however they surely show a betterment and similarity with the clean data results.

The acquisition performed on the filter data shows a lot of retracement towards the results of the original i.e. clean data as seen in figure 4c. Minor lobes on either side of the main lobe appear in the graph which had almost disappeared in the noisy data acquisition. Figure 4d shows tracking results of filtered data indicating improvement as compared to noisy data results. Similarly, navigational data in figure 4d also shows better results as compared to noisy data.

The s-curves for the different sets of navigational data such clean, noisy or filtered is illustrated in the figure 5. S-curve helps in determining the delay present in the signal. A comparison of s-curves for different types of navigational data helps in determining the amount of delay present in each data set. It is remarkable to see that despite having issues with noisy data the filtered data has attained the similar curve as the noisy one.

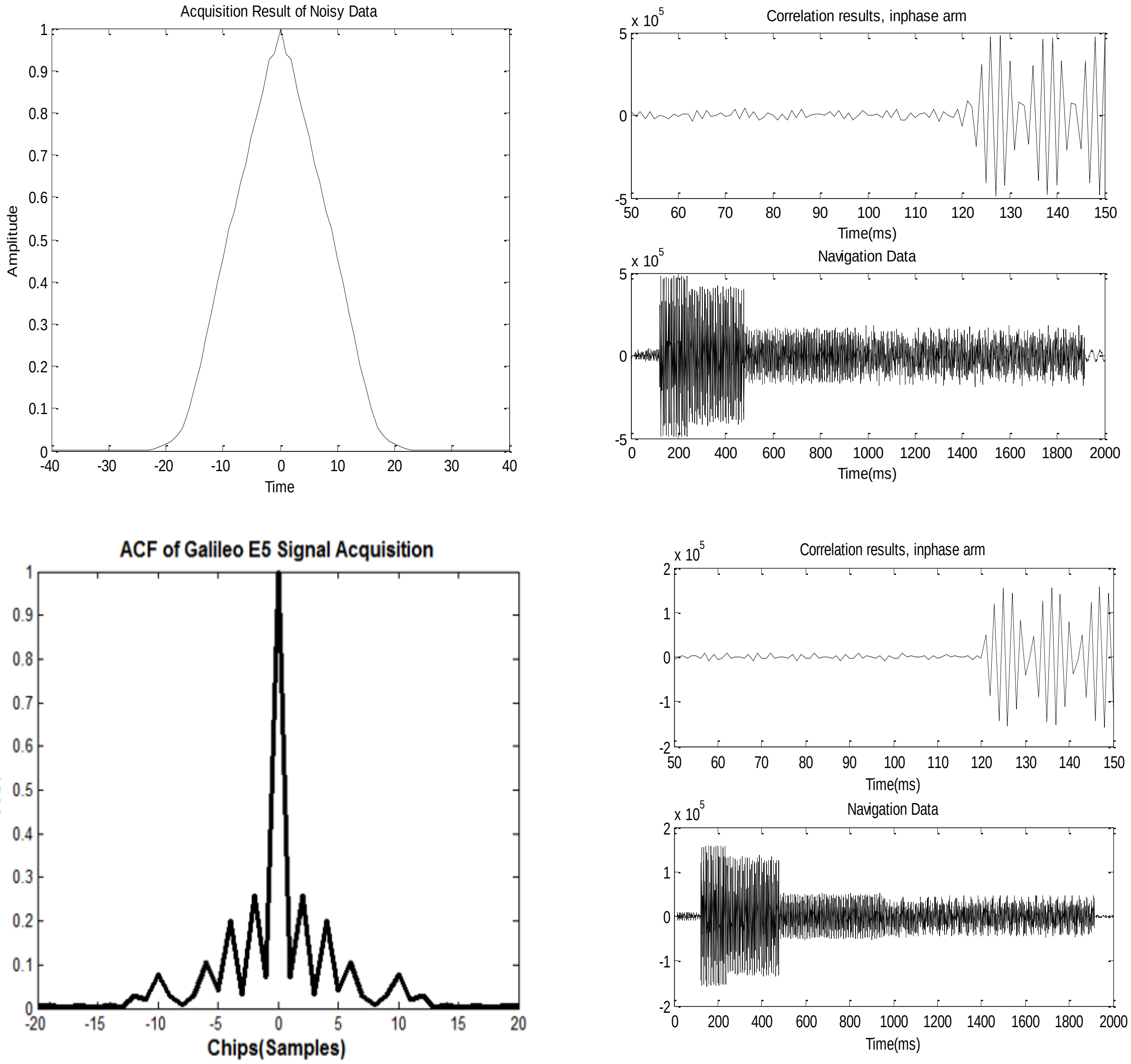

Figure 4: (a) Auto-correlation function of the signal acquisition under noisy dataset. (b) Correlation results from the signal tracking under the noisy datasets (c) Auto-correlation function of the signal acquisition after filtered datasets (d) Correlation results from the signal tracking after filtered datasets

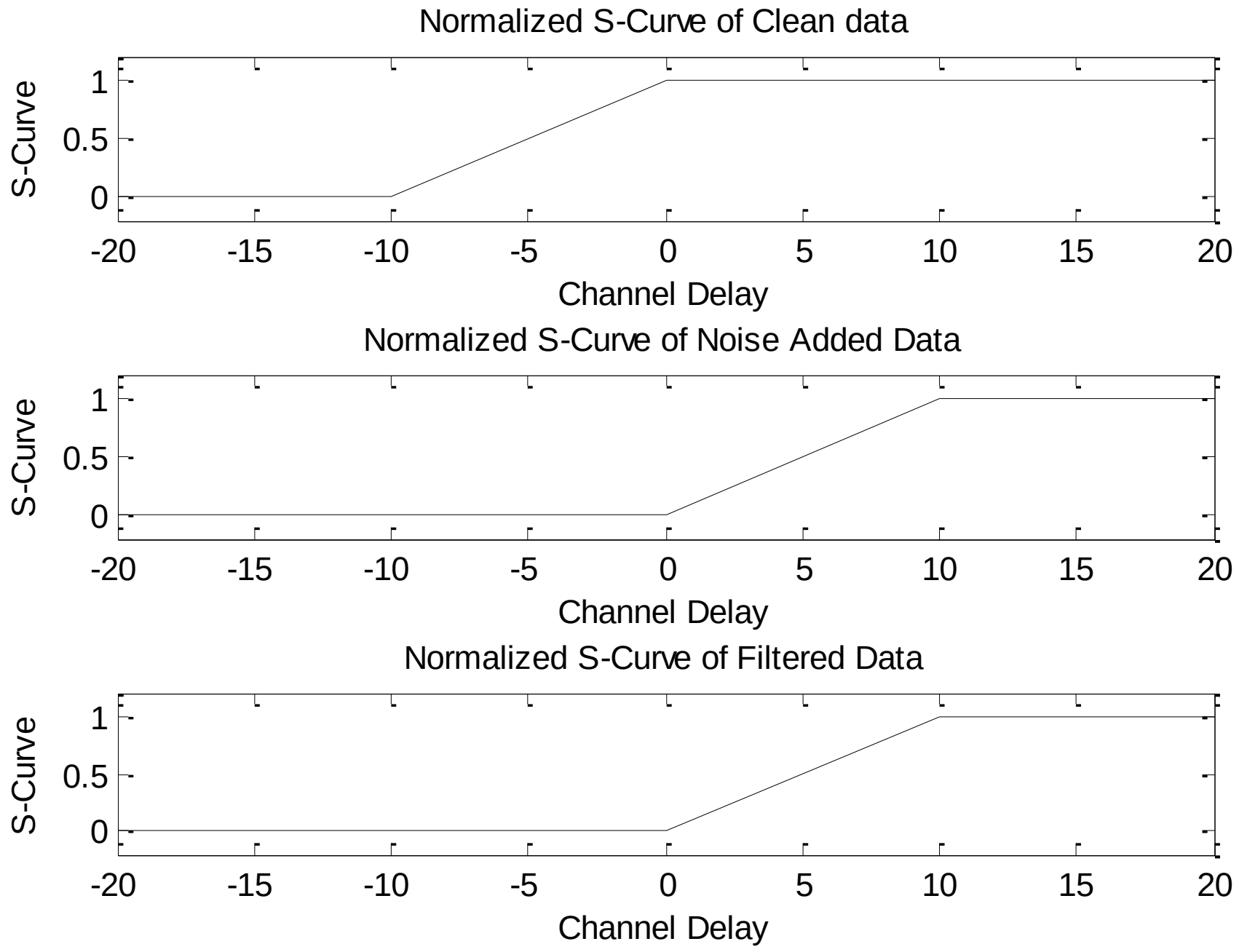

Figure 5: S-curve comparison of clean, noisy, and filtered datasets

# 6. CONCLUSION

This paper has discussed the unique learning of Galileo E5 AltBOC signal in the form of a comparison of clean, noisy, and filtered datasets. The entire baseband signal processing is performed in one single standalone software application to provide feasibility in testing the modulation scheme for Galileo E5 signal. The results obtained from the clean datasets have provided optimal results in the ACF response of acquisition and correlation in phase arm of the tracking channels to extract navigation data. However, the comparison of filtered and noisy datasets has provided sufficient information in distinguishing the ACF response of the acquisition and extraction of the navigation data. The positioning under all the three condition is then compared with an S-curve which shows the normalization of clean data in smoother way while keeping both filtered and noisy datasets in same plane.